# I. Introduction

The timing, success, and overall productivity of agricultural systems are significantly impacted by seed dormancy, a natural state that prevents a seed from germinating even under favorable conditions[1]. This survival strategy allows seeds to bypass unfavorable periods (like cold winters or dry seasons) and germinate when the chances of seedling survival are the highest. Dormancy can be due to several reasons like hard seed coats preventing water absorption, the presence of chemical inhibitors, or physiological blockage within the embryo[2]. In the case of sunflower seeds, two types of dormancy can be distinguished: (i) the integumentary dormancy where inhibitory mechanisms of the integument and pericarp prevent germination at temperatures above 25°C and (ii) embryo dormancy which inhibits germination at temperatures below 15°C[3]. In the case of non-dormant sunflower seeds, optimal germination is between 20 and 25 °C.

In addition to seed dormancy, the overall productivity of agrosystems is being significantly challenged due to the ever-increasing global population but also to the profound and enduring effects of climate change[4], [5], [6]. This dual pressure translates into considerable constraints on agricultural yields, compounded by exacerbating factors such as urbanization and loss of agricultural land, deforestation, increase in pests and diseases, sea-level rise and salinization, decreased genetic diversity and socioeconomic inequality[7], [8], [9].

Among these global factors, water pressure plays a pivotal role, taking the form of drought due to insufficient rainfall, high evaporation rates or inefficient soil moisture retention. In sunflower seeds, drought can lead to increased accumulation of linoleic acid and to significant decrease in oleic acid content accompanied with a reduction in oil yield[10], [11]. Conversely, water-saturated scenarios can lead to a shortage of oxygen, hampering seed germination and the early stages of growth[12]. In addition, irregular rainfall can create alternating periods of water abundance and deficiency, posing problems for young seedlings. On larger scales, climatic variations can amplify the frequency and intensity of droughts or floods, disrupting the regularity or volume of rainfall and increasing soil salinity due to rising sea levels or diminishing water supplies[13], [14], [15]. These water stress conditions can exacerbate the effects of dormancy on seed germination, as documented for several decades[16], [17].

The compounding effects of the aforementioned factors – whether seed-specific or global – highlight the pressing necessity for inventive and robust practices to enhance germination capacities and therefore securing global food supply in the years ahead. In response to these challenges and specifically drought issues, a range of seed treatment techniques are being employed today like breeding, biostimulants, priming and coating. Although not resorting increased pesticide usage, these techniques suffer from other limitations:

- Breeding and varietal selection are particularly relevant for large-scale crop seeds such as wheat, sunflower or rice to obtain the varieties showing optimal stress tolerance index[18], [19], [20]. However, these approaches are time-intensive and expensive, involving careful breeding and testing over several generations of plants[21]. Besides, the outcomes of breeding programs can be unpredictable, with desirable traits sometimes linked to undesirable ones, making it challenging to produce an "ideal" variety[22].
- Biostimulants suffer from a lack of standardized regulations, leading to discrepancies in quality and efficacy among products[23]. The effectiveness of biostimulants can vary greatly depending on the environmental conditions and the specific crop, leading to inconsistent results[24]. Besides, combinations of biostimulants and times of application can have very different results as shown by de Morais *et al.* for soybean seeds in water deficit conditions[25]. Moreover, high-quality biostimulants can be expensive, which may limit their use, especially in regions where the economic return might not justify the investment.
- Seed priming can be employed to initiate the physiological processes leading to germination in water scarcity conditions. Hence, sodium hydrosulphide (NaHS) solutions can be used to prime sunflower seeds (*Helianthus annuus L.*) exposed to different drought levels although providing mitigate results [26]. Overall, hydropriming (simple rehydration), osmopriming (osmotic treatment) and hormopriming (hormonal treatment) can be effective techniques but operate over long durations, ranging from several hours to several days[27].
- Biopolymer-based seed coating can be developed to increase germination and water-stress tolerance in semi-arid and sandy soils[28]. However, these coatings may not always be biodegradable in the context of solving drought stress, especially when superabsorbent polymers (potassium polyacrylate, polyacrylamide-based polymer) [29]. Last and not least, pelleting and coating technologies can also be expensive[30].

To overcome these limitations and provide new and innovative solutions, alternative technologies are therefore expected such as cold atmospheric plasma (CAP) processes. CAPs are ionized gases that exist at or near room temperature, created by applying energy to a gas, causing the gas particles to become ionized. Plasma, which is often described as the fourth state of matter, can exist either as a fully ionized gas, termed 'hot plasma' where there is thermodynamic equilibrium, or as a partially ionized gas, known as 'cold plasma.' In the latter, the presence of neutral species is significantly higher compared to the charged ones, and the electron temperature is notably higher than both the ion and neutral temperatures[31]. Cold plasmas are a versatile approach widely used to meet challenges in life sciences (medicine, environment, agriculture) especially thanks to their ability to generate short- and long-lived reactive species including O, OH, $NO/NO_2$, $O_2(a^1\Delta g)$, $O_3$, electronically excited $N_2$ ($N_2^*$), and metastables, in addition to their radiative, electrical, thermal and fluid-mechanical properties[32], [33].

The use of cold plasmas has attracted increasing attention in agricultural circles due to its proven benefits in improving seed germination characteristics, including seed vigor, dormancy release and promotion of seedling growth. These beneficial effects have been successfully demonstrated on a range of economically important crops, including tomatoes[34] and soybeans[35]. Cold plasmas exert their beneficial influence by modifying the physico-chemical characteristics of the seed surface[36]. For example, they improve surface wettability, a vital characteristic for successful imbibition and initiation of germination[37]. More recent research highlights the effectiveness of cold plasmas in combating phytopathogenic attacks on crops such as maize[38], barley and wheat[39], as well as in the degradation of mycotoxins[40]. These observations establish cold plasma as a versatile and promising tool for seed treatment to improve or maintain germination capabilities.

Several research works have already employed cold plasma technology to improve the germinative properties of seeds in water stress conditions. Under drought stress, Ling et al. have demonstrated that cold plasma treatment could slightly improve the germination rate of drought-sensitive and drought-tolerant oilseed rape seeds[41]. Seedling growth parameters (shoot and root dry weights, shoot and root lengths) also significantly increased after plasma exposure. Feng et al observed similar effects on alfalfa seeds after exposure to helium-air plasma (DBD) [42]. Under increasing drought stress conditions, the treated seeds showed vigor indexes I increased compared to their respective controls. Phenotypic observations have also been performed by Adhikari et al. on tomato seedlings under drought stress[43]. They demonstrate that seedlings obtained from seeds treated 10 min by an air plasma jet had higher total chlorophyll content. While many studies propose to use CAP to improve germinative properties of seeds under drought conditions (eg in wheat and cotton[44], [45]) only few has been specifically focused on seed dormancy. In Arabidopsis, however, August et al. (2023)[46] demonstrated that a short CAP treatment could significantly release dormancy for seeds with higher water content (i.e. > 10%$_{DW}$).

The objective of the present study was to investigate the effect of ambient air plasma on dormancy breaking and early seedling development in sunflower. The potential of this plasma treatment was tested under conditions of increasing water stress. To this end, two distinct experimental procedures were formulated: one involved the use of polyethylene glycol (PEG) solutions for seeds sown on cotton substrates housed in Petri dishes, while the other managed the water supply for seeds and, subsequently, seedlings, rooted in soil substrates contained in pots.

## 2. Material and methods

### 2.1. Experimental plasma setup

Ambient air plasma was generated in a dielectric barrier device (DBD) consisting of two planar electrodes of 30 × 40 cm of area, one connected to the high voltage and the other grounded. As sketched in **Figure 1**, the two electrodes were separated by an insulating dielectric material (classical soda-lime glass) 2 mm thick to prevent any streamer-to-arc transition and therefore ensure a homogeneous distribution of plasma in the reactor volume where the seeds were exposed to ambient air plasma. Here, the background gas was the ambient air with a relative humidity close to 40%. The high voltage electrode was supplied with 9 kV in amplitude at a frequency of 140 Hz (sine form) using a high voltage generator composed of a function generator (ELC Annecy France, GF467AF) and a power amplifier (Crest Audio, 5500W, CC5500).

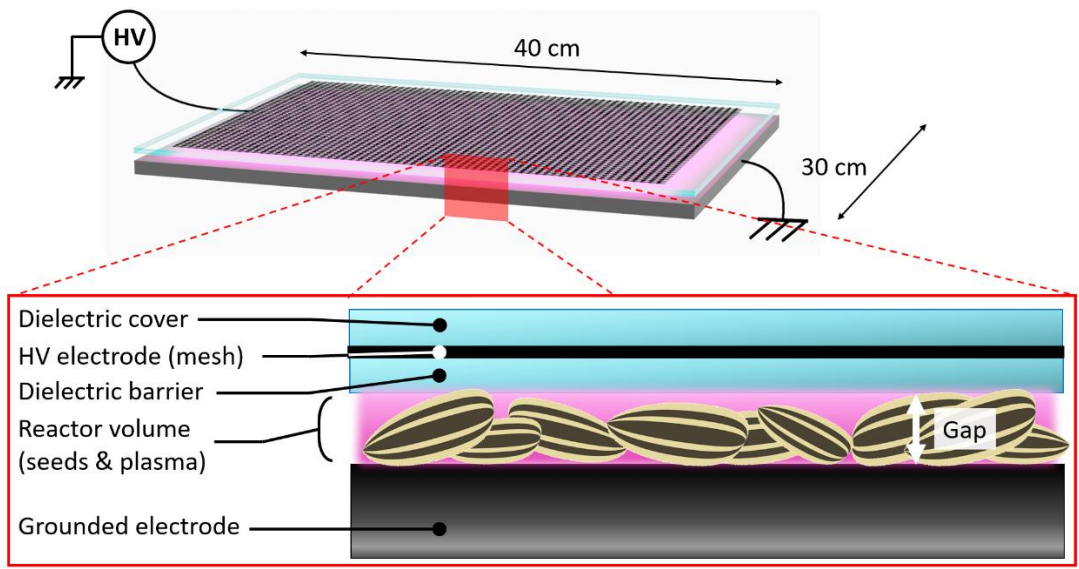

*Figure 1: Schematic of the dielectric barrier device operating in ambient air. Sunflower seeds are placed in the gap region 4 mm thick.*

## 2.2. Plasma diagnostics

Measurements of electrical parameters were conducted using a Teledyne LeCroy Wavesurfer 3054 digital oscilloscope in conjunction with a high voltage probe (specifically, Tektronix P6015A 1000:1, Teledyne LeCroy PPE 20 kV 1000:1, Teledyne LeCroy PP020 10:1), as well as a Pearson 2877 current monitor. The distributions of current peaks were derived from original current signals, with a temporal precision of 500 ns, following a tri-step methodology: (i) the dielectric component was eliminated using the peak analyzer toolbox of the Origin software, specifically utilizing the asymmetric Least Squares Smoothing procedure, (ii) signal noise, which is indicative of peak magnitudes less than 2 mA, was isolated, (iii) the local peaks were organized in a sequence from highest to lowest.

Optical emission spectroscopy was utilized to identify the radiative species of the plasma phase. The spectrometer (SR-750-B1-R model from Andor company) was equipped with an ICCD camera (Istar model) which operates in the Czerny Turner configuration. Its focal length was 750 mm while diffraction was achieved with a 1200 grooves.mm$^{-1}$ grating blazed at 300 nm. Light was collected with an ANDOR SR-OPT-8014 single-fiber bundle (100 µm core, HOH UV/VIS silica, 2 m length) coupled to the spectrometer entrance slit. The fiber tip was positioned 1 cm from the discharge, at 90° to the electrode plane, viewing the center of the gap. The spectrometer slit width was 80 µm. The spectral resolution FWHM was 15 pm at 337.1 nm, determined with a Hg/Ar calibration lamp; wavelength accuracy was ±0.05 nm. The following parameters were selected for all experiments: exposure time = 0.1 s, number of accumulations = 20, readout rate = 3 MHz, gate mode = 'CW on', Gain level = 3000, insertion delay = 'normal'.

A quadrupole-based mass spectrometer (Model HPR-20 from Hiden Analytical Ltd) was utilized to identify the gaseous species considering their m/Z ratio. They were collected by a quartz capillary, 1 m in length, flexible, chemically inert and heated at 200 °C to prevent chemisorption. Then, a three-stage differentially pumped inlet system separated by aligned skimmer cones and turbo molecular pump, enabled a pressure gradient from $10^5$ bar to $10^{-7}$ bar at the entrance of the ionization chamber. There, ionization energy was set at 70 eV. The residual gas analyzer (RGA) mode was used for scanning masses from 1 to 50 amu, using a secondary electron multiplier (SEM) at 1600 V.

## 2.3. Plant material

This study was performed with seeds of two commercial sunflower (*Helianthus annuus* L.) hybrids, referred here as variety A and variety B, produced by NuSeed (USA) in 2020. Seeds were stored at −18 °C after harvest in order to maintain their initial physiological state[47]. At harvest, the two varieties exhibited distinct germination kinetics: variety A showed slower germination indicative of a dormant-like behavior, while variety B germinated rapidly at 15°C. These contrasting germination profiles were used to investigate the effects of treatments under different physiological conditions.

## 2.4. Germination assays

Seed dormancy was assayed by distributing whole achenes into three separate replicates, each consisting of 25 seeds, and placing them onto a layer of cotton saturated with deionized water within 9 cm diameter glass Petri dishes at 15°C[47] Germination was scored daily for 15 days, a seed being considered as germinated when the radicle protruded through the pericarp. The results presented are means ± SEM of germination percentages obtained with 3 replicates. Results were also expressed as areas under the curve (AUC) which refers to the cumulative germination curve and which plots germination percentage (or number of seeds germinated) over time.

The Area Under the Curve (AUC) was calculated using the trapezoidal rule as follows:

$$AUC = \sum_{i=1}^{n-1} \frac{(y_i + y_{i+1})}{2} \cdot (x_{i+1} - x_i)$$

Where $x_i$ represents time (in days), and $y_i$ corresponds to the percentage of germination at time $x_i$.

To decipher the effects of water stress on seed germination, achenes were germinated on a range of polyethylene glycol (PEG) solutions giving water potential values from −0.4 to −1.0 MPa[48]. Germination results in the presence of PEG were analysed using the hydrotime model[49] which allowed the calculation of $\Psi_{50}$, the water potential at which 50% of the seed population germinated, providing a comparative measure of drought sensitivity. Water uptake was calculated over a 23-hour period for seeds of variety B placed either on water or on a solution of PEG -0.8 MPa, using the following formula:

$$Water\ uptake\ (\%) = \frac{W_t - W_0}{W_0} \cdot 100$$

Where $W_0$ is the initial dry weight of the seeds and $W_t$ is the seed weight after 23 hours of imbibition.

## 2.5. Greenhouse experiments

In greenhouse experiments with seedlings, an initial process consisted in dehydrating 100 g of Jiffy Substrates V1 potting soil at 105 °C until its relative humidity reached 0%. The soil's maximum retention capacity (100% humidity) was then determined by pouring a measured amount of water onto the dehydrated soil, which was placed in a filter within a funnel atop a graduated cylinder. The difference between the initial volume of water and the volume recovered in the cylinder provided the maximum amount of water the soil could retain. This data served to calculate the precise amount of water required to achieve specific soil moisture percentages. This preliminary procedure ensured the possibility of maintaining controlled watering, thus preserving a steady and specific humidity percentage throughout greenhouse growth. The same treatment was applied to all potting soil used, enabling the preservation of initial watering conditions. Necessary adjustments were made based on daily weight assessments.

Three sets of seeds, later grown into seedlings, were used for greenhouse experiment: control seeds (C: unexposed to plasma) and seeds subjected to 15 ($P_1$) and 30 min ($P_2$) ambient air plasma exposure. In each case, 20 seeds were sown in pots at a density of 4 seeds per pot. Two distinct humidity conditions were adopted for growth: one ensured soil moisture content exceeding 70% within the pot (standard condition) and the other maintained 40% soil moisture content (stress condition). After initially fixing the soil's water saturation percentage, pots were weighted daily and watered with the appropriate water volume to maintain the initial soil moisture conditions. To mitigate potential biases such as border effects, seeds from various treatment groups were randomly allocated on the cultivation trays and repositioned daily. These trays were then housed in a greenhouse with a consistent temperature of 21°C. Following emergence from the soil, the length of the sunflower hypocotyls was recorded each day to monitor seedling growth progression.

In order to integrate both germination and early growth parameters, the vigor index I was calculated for each condition by multiplying the mean hypocotyl length (cm) of germinated seedlings by the corresponding germination percentage, and dividing the result by 100, according to the following formula:

$$Vigor\ index\ I = \frac{\left(\begin{array}{c}Mean\ Hypocotyl\\ Length\ (cm)\end{array}\right) \times \left(\begin{array}{c}Germination\\ Percentage\ (\%)\end{array}\right)}{100}$$

# 3. Results

## 3.1. Characterization of the gaseous plasma phase

The electrical characteristics of ambient air plasma were examined with a focus on two factors: first, the impact of the gap distance, and second, the presence or absence of seeds within the reactor volume. The determination of the optimal gap value is particularly significant as it results from a compromise between:
- Seed biology (the size of the seeds, and specifically the lowest dimension of the seeds which, in the case of sunflower, is estimated to $D_{low}$ = 3.8 mm)
- Plasma physics (the voltage breakdown is ruled by the Paschen law while operation voltage must ensure a continuous operation of ambient air plasma out of thermodynamic equilibrium).

Whatever the gap value (1, 4 or 7 mm) with/without seeds in the reactor volume, the **Figure 2a** indicates that the plasma voltage remained unchanged, with an amplitude as high as 9 kV and a frequency of 140 Hz. In that respect, two issues should be highlighted:
- Regarding the applied voltage, it is worth stressing that it was fixed by the high-voltage amplifier at 9 kV (peak) for all gaps. Therefore the applied voltage waveform was identical in **Figure 2a**. The gap primarily affected the discharge current pulse statistics and the plasma power, consistent with equivalent-circuit descriptions of DBDs [31]. Conversely, the time profile of the current peaks changed drastically considering different gap values, as shown in **Figure 2b** (without seeds). While 40 mA current peaks were obtained for a 1 mm gap, the current peak amplitude distribution was drastically reduced to values lower than 2 mA at 4 mm and 7 mm. Seeds could be introduced in the reactor volume only for gaps equal to at least 4 mm. As a result, the overall distribution of the current peak amplitude was significantly increased in **Figure 2c** compared with **Figure 2b**.

- Regarding the operating frequency, we selected the value of 140 Hz because it ensures efficient power transfer while maintaining a cold, spatially uniform discharge. In low-frequency operation, each half-cycle (about 3.6 ms at 140 Hz) allows surface charges on the dielectric to redistribute, so microdischarges ignite at different positions over the electrode area. This results in a diffuse glow, consistent with large-area homogeneous treatment. At higher frequencies (>300 Hz), current pulses overlap in time, charge relaxation is incomplete, and streamers tend to re-ignite at the same locations, producing bright filaments and non-uniformity[46]. In addition, the slower rise of the applied voltage at 140 Hz enables clear resolution of individual current peaks, which is essential for the statistical analysis shown in **Figure 3**.

To more effectively evaluate the influence of the gap on the distribution of current peak amplitudes, each time-dependent signal was observed over the span of seven periods, utilizing three distinct triplicate sets. As outlined in the methodology (Section 2.2), the distribution of current peak amplitudes can be graphically represented on a log-log scale, as demonstrated in **Figure 3a**. For a gap distance of 1 mm (illustrated by the black curve), around 10 000 peaks were discerned, with a subset of 10 of these exhibiting magnitudes surpassing 30 mA, and a larger subset of approximately 9 000 streamers demonstrating amplitudes in the range of 2 to 7 mA. When the gap was extended to 1.5, 2, and 2.5 mm, there was a corresponding increase in the quantity of peaks possessing high magnitudes. Nonetheless, this tendency was inverted for larger gaps. The curve corresponding to a 3 mm gap presented an intriguing profile, notable for its high reproducibility, signifying a substantial decrease in low-amplitude peaks. This occurred because the gap enlarged to such an extent that it inhibited the corresponding micro-discharges from bridging the dielectric barrier with the counter-electrode, thereby limiting the operational discharges to a coarse estimate of a hundred (within the 10-400 mA range). As the gap continued to increase, a diminishing quantity of current peaks, typically less than 100, were still observed, presenting with relatively low current magnitudes (< 20 mA). This suggests isolated and stochastic electrical events that could no longer be associated with a homogeneous plasma discharge.

In the seed-packed configuration, microdischarges occur within a noticeably narrower time window around each voltage crest (**Figure 2c** versus **Figure 2b**). Physically, the seeds act as a distributed dielectric 'bed': once ignition begins, many seed asperities experience field enhancement simultaneously. The ensuing streamers deposit surface charge over a very large effective area (seeds + barrier), which rapidly opposes the applied field so that the gas field $E_{gas}$ falls below breakdown soon after the crest. In parallel, hydrated pericarps are slightly conductive, shortening the effective RC time of the gap. Therefore, charges redistribute and neutralize more quickly than in the empty reactor. These two effects (phase-synchronous ignition at enhanced-field sites and fast charge accumulation/relaxation) both compress the active portion of each half-cycle. The result is a higher instantaneous pulse rate but a shorter overall 'on-time': pulses are more numerous, individually narrower, and strongly bunched near the crest, with little activity on the rising/falling shoulders. This temporal gating explains the tighter clusters seen in **Figure 2c**.

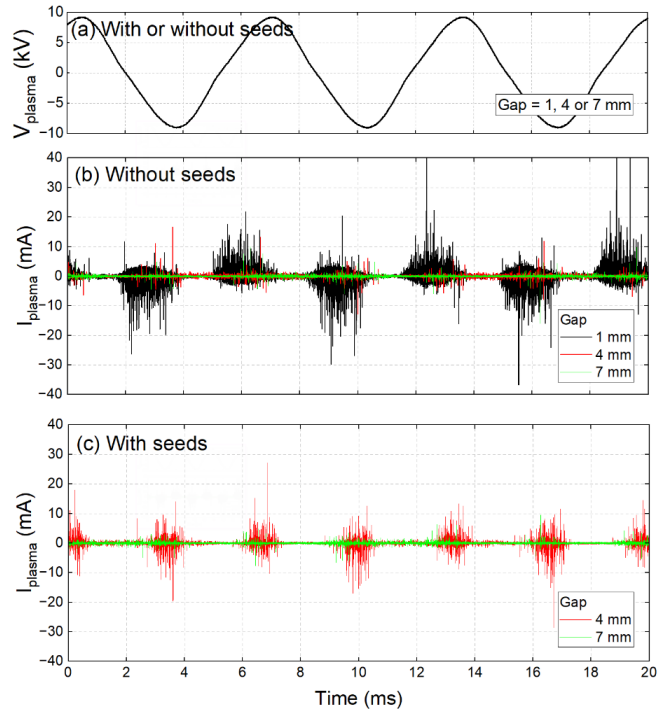

*Figure 2: (a) Time profile of the plasma voltage measured for gaps of 1, 4 and 7 mm without seeds and with sunflower seeds (20 g) in the reactor volume, (b) Time profile of the plasma current without seeds in the reactor volume for gaps of 1, 4 and 7 mm, (c) Time profile of the plasma current with seeds in the reactor volume for gaps of 4 and 7 mm.*

If the gap distance has an influence on the current peak amplitude distribution without seeds in the DBD, it is then important to study the same phenomenon when seeds are introduced in the reactor volume. For the sake of clarity, **Figure 3b** was only focused on minimum gap value at which seeds can be inserted in the volume reactor (gap = 4mm) and the gap value beyond which not even a micro-discharge can bridge the dielectric barrier with the counter-electrode (gap = 7 mm). At 4 mm, the introduction of sunflower seeds permitted a strengthening of the current peak amplitude distribution with a number of micro-discharges increasing from 200 (without seeds) to 2000 (with seeds) and maximum peak amplitudes increasing from 20 to 32 mA respectively. At this gap value, the micro-discharges could bridge the dielectric barrier with the counter electrode but also the dielectric barrier with the seed. For the larger gap at 7 mm, the same behavior was observed although the current peaks were so low in amplitude (< 10 mA) and so few in number (< 100) that they were unlikely to induce the expected biological effects.

While a gap of 4 mm seems appropriate according to the current peak amplitude distributions, another important parameter to assess is the electrical power deposited by the plasma in the reactor volume. This parameter was measured by the Lissajous method which required placing a measurement capacitor between the counter-electrode and the ground. The value of the capacitance (100 nF) was chosen at least 10 times higher than the one of the dielectric barrier (3.7 nF) to be non-intrusive[46]. Then, it was possible to plot the plasma voltage ($V_{plasma}$) as a function of the plasma charge ($Q_{plasma}$), as proposed in **Figure 3c** for a gap of 4 mm. Multiplying the area of the closed contour by the applied frequency gave the value of the plasma power, which was estimated to 430 mW and 2.5 W without seeds and with seeds in the reactor volume respectively.

A more general view of this discrepancy is reported in **Figure 3d** where the plasma power is plotted versus the gap, considering a reactor volume without seeds (black curve) and with seeds (red curve). In the first case, a slight increase in the power was obtained between 1 and 1.5 mm of gap, from 30.8 to 36.4 W before decreasing for larger gap values. Interestingly, a sharp fall of the plasma power (from 36.4 to 560 mW) was observed between 1.5 and 3.5 mm, i.e. on a very narrow gap range. Owing to the $D_{low}$ value of 3.8 mm, the sunflower seeds could only be introduced for a gap of at least 4 mm. In that case, the red curve indicates that the plasma power was significantly increased from 430 mW to approximately 2.8 W. For increasing gap values, the plasma power always decreased to lower values and remained higher when seeds were present in the reactor. These results indicate that an ambient air plasma generated in a dielectric barrier device may be a limitative option to treat seeds of larger dimensions and that the gap was a crucial parameter requiring a fine tuning.

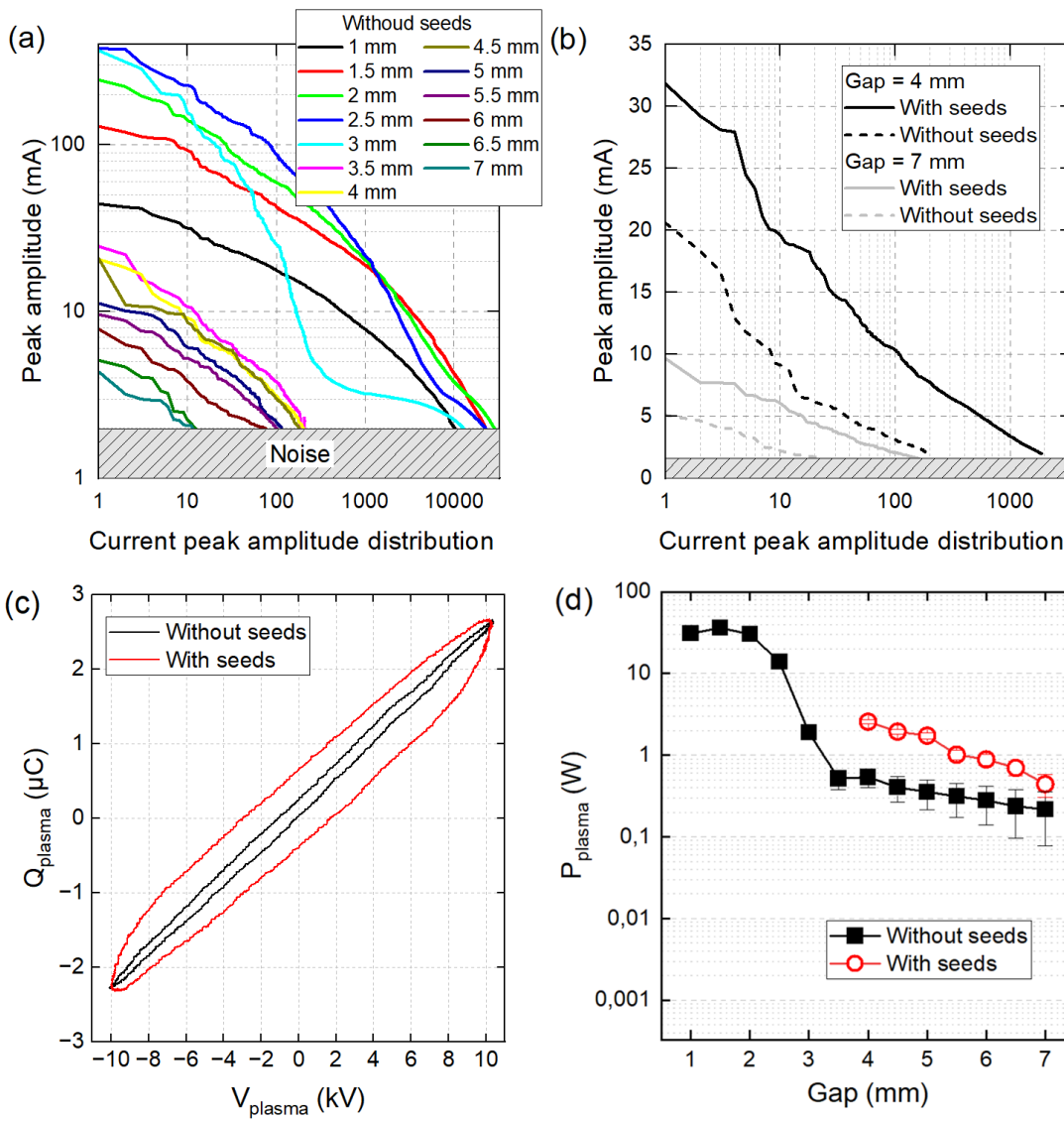

*Figure 3: (a) Current magnitude distribution of current peaks in ambient air plasma for different gaps without seeds, (b) Same distribution for gaps of 4 and 7 mm, with or without seeds, (c) Lissajous curves at 4 mm gap without and with seeds (d) Plasma power measured by Lissajous method as a function of the gap without/with seeds in the reactor volume of the DBD. Each measurement was performed in quadruplicate; error bars represent the standard error of the mean (SEM).*

The importance of gap settings is underlined by the electrical characterization and can be complemented by the photographic report in **Figure 4**. The reactor volume in the DBD was captured from a side view, taking into account different scenarios: (i) plasma off versus plasma on, (ii) without seeds versus with seeds, (iii) gap distances ranging from 1 mm to 7 mm. Several observations are noteworthy:

- When the gap was only 1 mm, the ambient air plasma could uniformly occupy the entire reactor volume. Then, it became more filamentary at 2 mm, strongly heterogenous at 3 mm and almost undetectable at 3.5 mm. Beyond 3.5 mm, it was almost imperceptible because the micro-discharges could occur randomly across the entire 30 × 40 cm$^2$ surface electrode, and did not necessarily appear in the area captured by the camera (limited to a side view of around 2 cm). In the absence of seeds and with larger gaps, no other micro-discharge was detectable.
- The inclusion of seeds in the reactor volume made the ambient air plasma more apparent when the gap was 4 mm. Interestingly, the plasma completely surrounded each seed, without forming a uniform layer filling the whole reactor volume. At 4.5 mm, the ambient air plasma still surrounded each seed although partially. At a gap of 5 mm, only one seed could interact with the plasma, and at larger gaps, no plasma was detectable.
- In the empty reactor, visible plasma emission is confined to gaps ≤3 mm and disappears beyond 3.5 mm. With seeds present, plasma reappears at the seed surfaces for gaps of 4-4.5 mm, confirming that field enhancement on the seed coat sustains localized micro-discharges. At 5 mm, emission becomes weak and patchy, and for ≥6 mm no discharge is detected. Thus, stable treatment is practically restricted to 4-4.5 mm, in agreement with the power trends of **Figure 3d**.

These results highlight that when treating seeds with ambient air plasma, an accurate adjusting of the gap value (precision of about 0.1 mm) is crucial to optimize the plasma-seed interaction. In particular, ambient air plasma might be a less recommended approach for treating seeds with excessively heterogeneous size distributions.

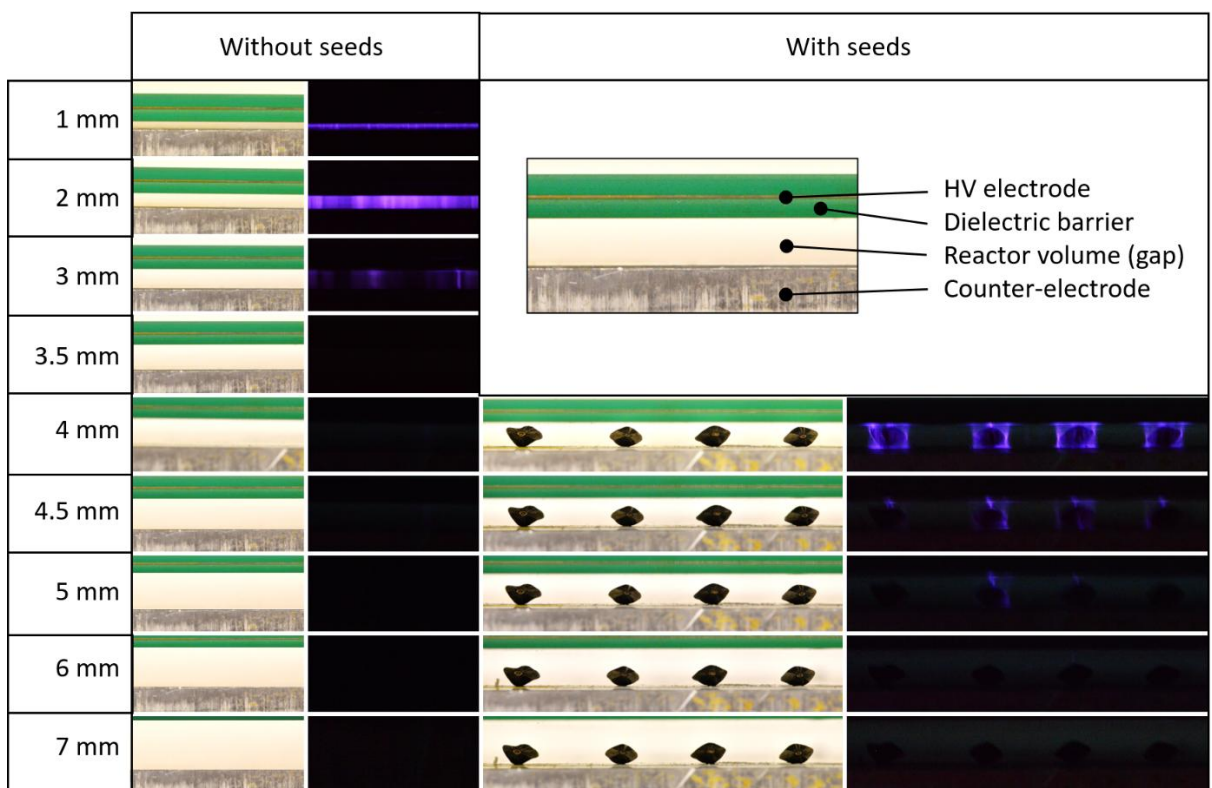

*Figure 4: Photographs of the DBD source viewed from the side, without seeds or with 4 seeds of sunflowers, without or with ambient air plasma (9 kV, f = 140 Hz), for various gap values. When room light is switched on, camera exposure time and aperture are 1/40 s and f/22 respectively. When room light is switched off, camera exposure time and aperture are 4 s and f/22 respectively. These later values are deliberately chosen to capture more detailed and clear images of the plasma. The case "with seeds" includes 4 aligned seeds for the sake of clarity.*

Although camera imaging provides insights into how the ambient air plasma fills the reactor volume and how this filling depends on gap adjustments, the exact nature of the excited gaseous species remains unknown. Therefore, optical emission spectroscopy (OES) was carried out on the 250-800 nm range, considering 1-7 mm gaps. The OES spectrum in **Figure 5a** revealed that the ambient air plasma was composed solely of molecular nitrogen bands from the second positive system (SPS), characterized by the $C^3\Pi_u$-$B^3\Pi_g$ electronic transition states. This spectrum was derived without seeds and at a gap of 1 mm. Packing the DBD with seeds or extending this gap up to 7 mm neither added nor removed any lines or bands, but it did affect the overall emission of the SPS. This change was readily apparent in **Figure 5b**: without seeds, the optical emission of the band at 337.1 nm increased from 4.3 × 10$^5$ (at 1.0 mm) to 7.1 × 10$^5$ a.u. (at 1.5 mm) before decreasing sharply and vanishing completely at 3.5 mm. When sunflower seeds were placed in the interelectrode region, the minimum accessible gap was 4 mm. Consistent with the imaging results in **Figure 4**, the N$_2$ emission intensity at this gap reached 6.4 × 10$^5$ a.u., whereas no signal was detected at the same gap in the empty reactor. At larger gaps the emission again decayed rapidly, with no photons detected beyond 5.0 mm. As expected, these trends closely mirror the dependence of plasma power on gap distance in **Figure 3d**.

SPS was detected because the high-energy electrons in the plasma phase predominantly excited nitrogen molecules to the C$^3\Pi_u$ electronic state, as shown in equation {1}. These electrons can deliver sufficient energy to induce vibrational excitation of the excited nitrogen molecules (N$_2$*) within the C$^3\Pi_u$ state, allowing them to further relax back to a vibrational level of the lower B$^3\Pi_g$ electronic state through the emission of a photon, as depicted in equation {2}. Additionally, vibrational relaxation can proceed non-radiatively through collisions with other molecules. This mechanism, called quenching, is in competition with radiative de-excitation, as seen in equation {3}. In all three equations, v' and v" represent the initial and final vibrational levels, j' and j" stand for the initial and final rotational levels, and M symbolizes another molecule.

N$_2$ + e → N$_2$*(C$^3\Pi_u$, v', j') + e    {1}
N$_2$*(C$^3\Pi_u$, v', j') → N$_2$ (B$^3\Pi_g$, v", j") + hv    {2}
N$_2$*(C$^3\Pi_u$, v',j') + M → N$_2$ (v", j") + M    {3}

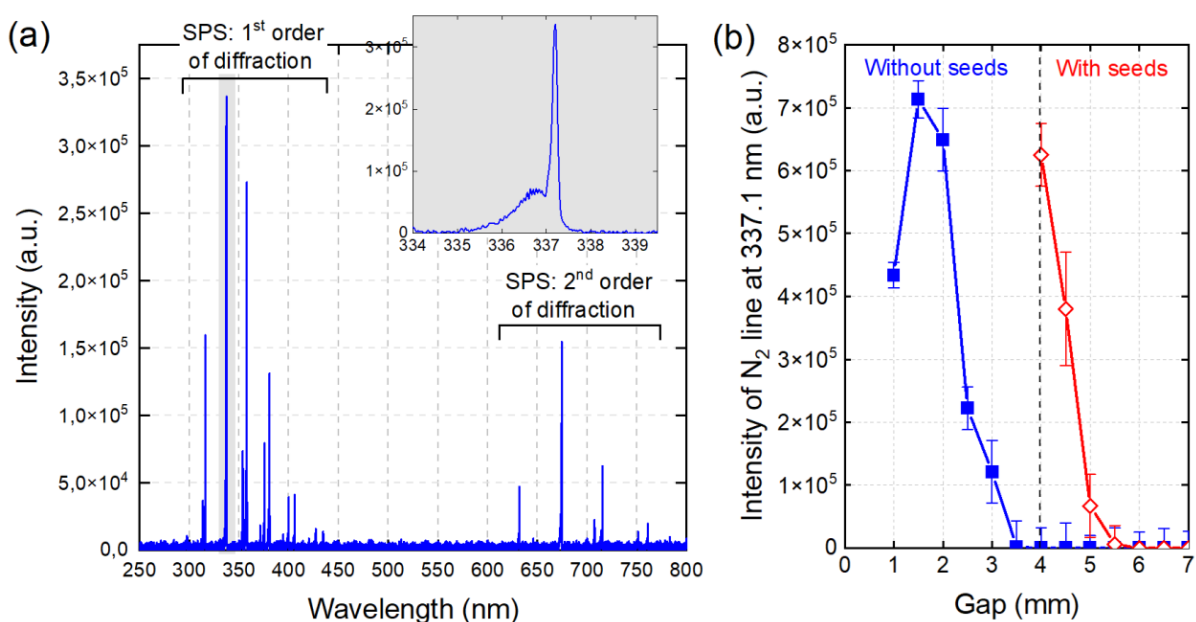

*Figure 5: (a) Optical emission spectrum of ambient air plasma measured in the reactor volume of the DBD, dominated by the second positive system (SPS) of molecular nitrogen (gap = 1 mm). Inset is a magnification view of the molecular nitrogen band headed at 337.1 nm. (b) Intensity of the N$_2$ band line at 337.1 nm as a function of gap. Each measurement was performed in quadruplicate; error bars represent the standard error of the mean (SEM).*

While the vibrational structure of SPS is always clearly visible in the OES spectrum, it is surprising that no gaseous oxygenated species was detected. It is particularly relevant to remind the following points :
- O radicals are commonly observed through the green line at 557.7 nm and the infrared lines at 777.4 nm, 844.6 nm and 926.6 nm. However, no such line was detected in **Figure 5a**.
- The electronic transition (A$^2\Sigma^+$-X$^2\Pi$) of the hydroxyl radical (OH) is often used owing to its strong emission at around 306-309 nm. The upper excited state (A$^2\Sigma^+$) is commonly populated through the dissociation of water vapor or electron-impact excitation of OH molecules. Once in the A$^2\Sigma^+$ state, the OH radicals can decay to various vibrational levels of the X$^2\Pi$ ground state, emitting photons. Although this transition is spin-allowed and has a relatively high Einstein A coefficient, the OH band was not detected in **Figure 5a**.
- The lowest excited states of O$_2$ are triplets, which are forbidden to decay radiatively to the singlet ground state. This results in a low probability for emission (a low "Einstein A coefficient") and long lifetimes for these states. Thus, the emission from O$_2$ is generally much weaker than from N$_2$ and is more challenging to observe, especially in an air plasma dominated by molecular nitrogen.

The inability of OES to quantify these oxygenated species could indicate their absence from the plasma phase, or alternatively, their existence in a quenched state. To elucidate this ambiguity, we undertook analytical investigations employing mass spectrometry. **Figure 6a** represents a mass spectrum (MS) of the ambient air plasma measured in a 1 mm gap, after air background correction (plasma off). Several noteworthy species could be identified considering their m/Z ratios expressed in g.mol$^{-1}$: N (14), O (16), OH (17), H$_2$O (18), N$_2$ (28), NO (30), O$_2$ (32), CO$_2$ (44) and O$_3$ (48). After acquiring several spectra without/with seeds for gaps between 1 and 7 mm, the line intensities of the previous species were measured when plasma was off and on so, as to derive ratios plotted versus gap, as shown in **Figure 6b** (without seed in the reactor volume). It turns out that all the short lifetime species (O, N, OH and NO) presented ratios approximating 1, thereby suggesting they were not produced within the ambient air plasma. The same inference could be made for long lifespan species with the exception of ozone which exhibited a line intensity peaking for a 2 mm gap. This line intensity ratio dipped to a value of 3 for a 4 mm gap when the reactor volume was devoid of seeds. Interestingly, the introduction of sunflower seeds changed this MS profile. **Figure 6c** illustrates that for a 4 mm gap, this ratio escalated to 15; a value corresponding to a 1.5 mm gap in **Figure 6b** (without seeds), and equivalent to the average distance partitioning the lower dielectric barrier surface from the uneven upper surface of the sunflower seeds.

be required, but these were currently unavailable in our laboratory. No additional species were detected by OES or MS when changing the gap from 1 to 7 mm, without/with seeds in the reactor volume. For the rest of the article, all plasma treatments applied to treat the sunflower seeds were conducted with a 4 mm gap.

## 3.2. Effect of ambient air plasma on seed germination

Seeds of varieties A and B were subjected to a 15 min ambient air plasma treatment using the Dielectric Barrier Discharge (DBD) method, as illustrated in **Figure 1**, and their germination was assessed at 15°C on water and on PEG solutions.

*Table 1. Water potential for 50% germination (Ψ50) at 15 °C. ND, not determined (germination too low).*

|  |  | Control | Plasma (15 min) |
|---|---|---|---|
| Batch | A seeds | ND | -0.57 MPa |
|  | B seeds | -1.15 MPa | -1.29 MPa |

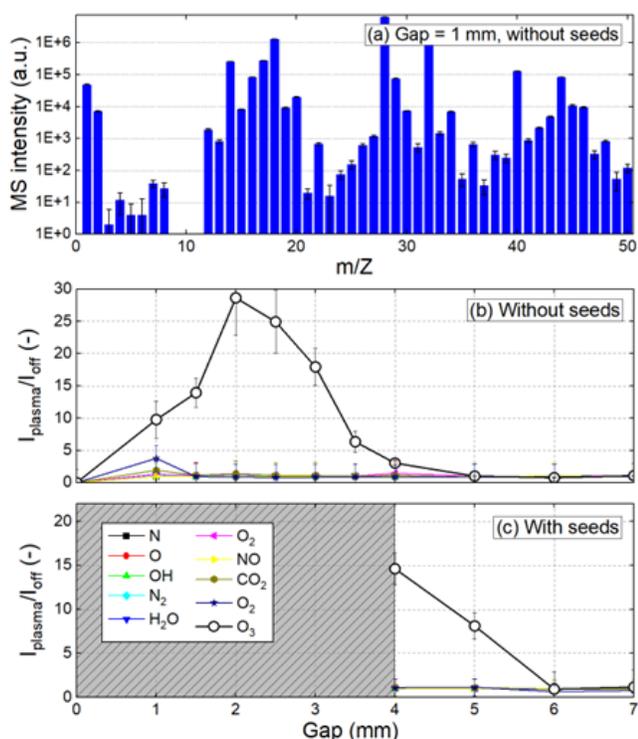

*Figure 6: (a) Mass spectrum of ambient air plasma generated in DBD for a gap of 1 mm, (b) Intensity ratio of the lines of each species detected by mass spectrometry as a function of the gap containing no seeds, (c) Same as (b) with seeds contained in the reactor volume. Each measurement was performed in quadruplicate; error bars represent the standard error of the mean (SEM).*

The active species detected in the ambient air plasma were molecular nitrogen excited in the SPS and ozone molecules. Short lifespan reactive species (OH, NO, O) could not be detected by OES because of quenching or by MS because of potential recombination into more stable compounds during propagation within the spectrometer capillary. Still, the existence of O radicals was expected to explain the production of ozone. To effectively detect these species, techniques such as laser-induced fluorescence (LIF) or cavity ring-down spectroscopy (CRDS) would

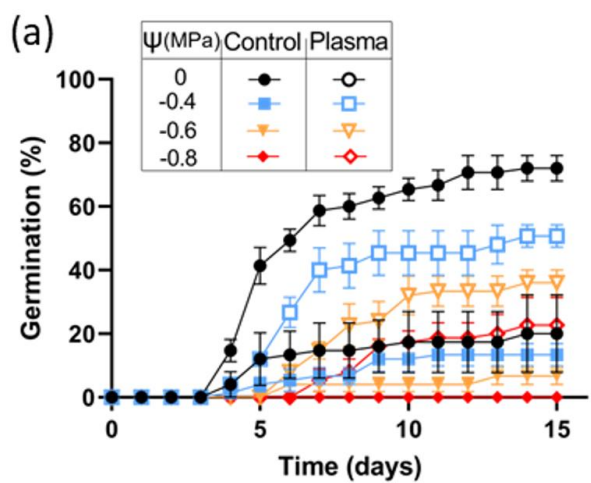

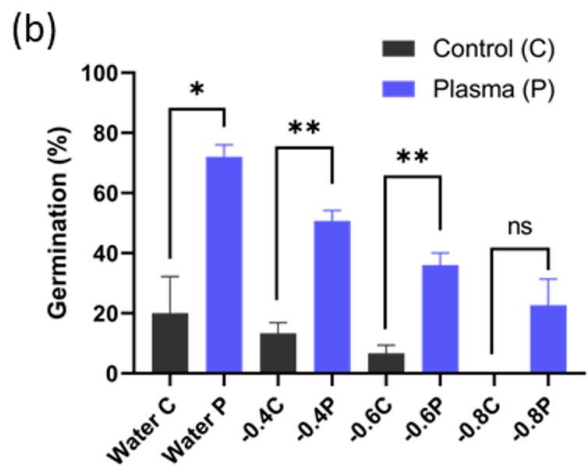

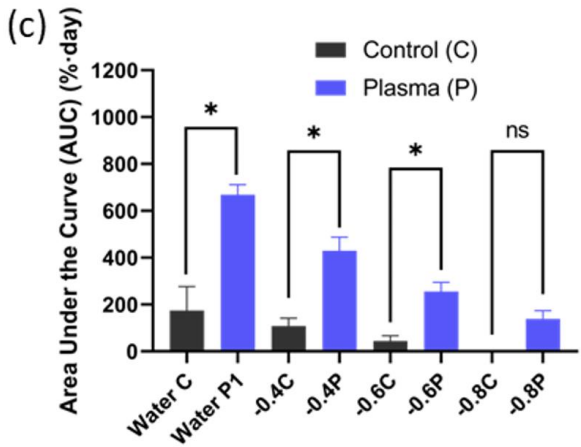

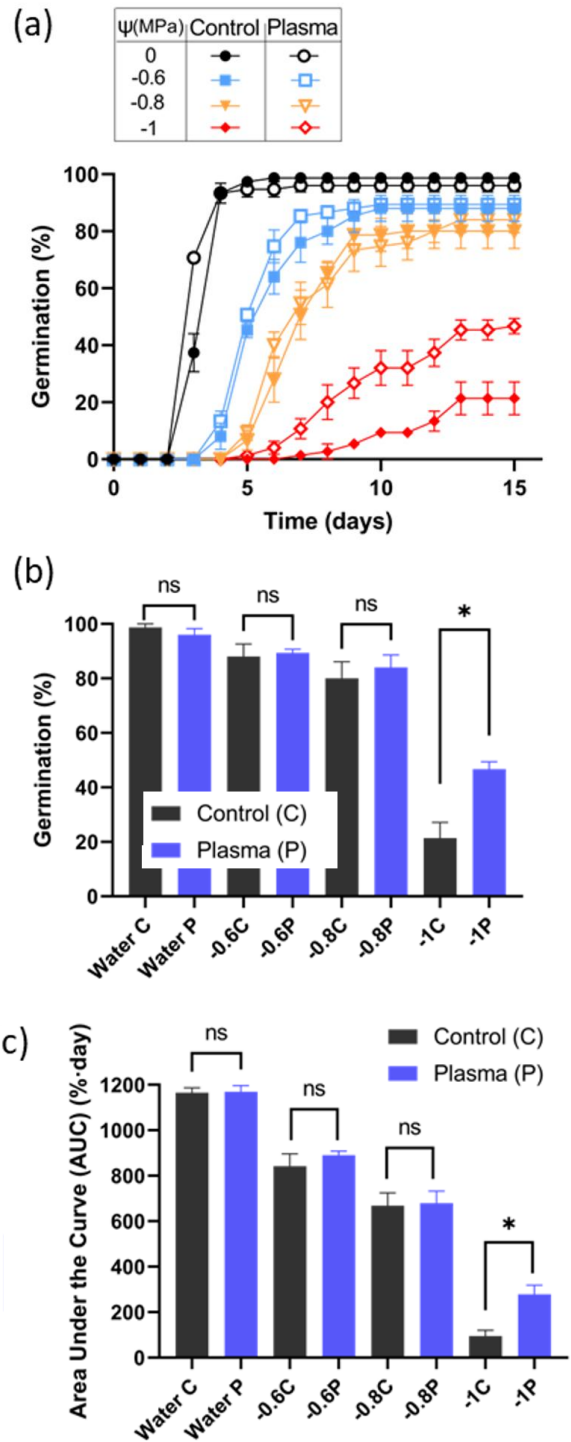

*Figure 7: Effect of plasma treatment on germination at 15 °C of sunflower seeds from variety A under PEG-simulated drought stress: Germination kinetics (a), final germination rate (b), and area under the curve (c). Each point or bar represents the mean of 3 biological replicates (25 seeds per replicate). Error bars indicate the standard error of the mean (SEM). Seeds were either untreated controls (C) or plasma-treated (P), and tested under no water stress (0) or increasing levels of PEG-induced water stress (–0.4, –0.6, and –0.8 MPa).*

Germination of A seeds on water did not exceed 20 % after 15 d at 15°C, suggesting that they were dormant (**Figure 7a, 7b**). In contrast, plasma-treated seeds exhibited a germination rate of about 70%, suggesting that the treatment partially released seed dormancy. As expected, B seeds fully germinated at 15°C on water within 4 days, but plasma treatment did not significantly improve their germination (**Figure 8a, 8b**). Increasing PEG concentrations in the imbibition media progressively reduced germination kinetics and final percentage of A and B control seeds (**Figure 7, Figure 8**). A seeds were very sensitive to water stress since their germination decreased at -0.4 MPa and was fully prevented at -0.8 MPa (**Figure 7a, 7b, 7c**). In contrast seeds from the B batch were much less sensitive to water stress since they germinated to 80 % at -0.8 MPa (**Figure 8a, 8b, 8c**). At -1.0 MPa however their germination did not exceed 20 % (**Figure 8a**). The positive effect of plasma on seed germination was clearly shown with dormant A seeds for which final germination percentage increased from 10 to 50 % at -0.4 MPa and from 5 to 35 % at -0.6 MPa (**Figure 7a, 7b**). For non-dormant B seeds, the effect of CAP treatment was less pronounced and became clearly evident only at -1.0 MPa, where water stress significantly reduced germination (**Figure 8a, 8b**). However, differences in the germination kinetics were already noticeable from -0.6 MPa. **Figure 7c**, where germination results are expressed as area under the curve, highlights the beneficial effect of CAP treatment on both final germination rate and mean germination time across all conditions, except under the most severe water stress. In contrast, in **Figure 8c**, the area under the curve of treated seeds is significantly higher only under the most severe water stress condition. Consistently, the decrease in $\Psi_{50}$ values following CAP treatment (**Table 1**) indicates that the treatment enhanced the ability of sunflower seeds to germinate under water stress conditions.

*Figure 8: Effect of plasma treatment on germination at 15 °C of sunflower seeds from variety B under PEG-simulated drought stress: Germination kinetics (a), final germination rate (b), and area under the curve (c). Each point or bar represents the mean of three biological replicates (25 seeds per replicate). Error bars indicate the standard error of the mean (SEM). Seeds were either untreated controls (C) or plasma-treated (P), and tested under no water stress (0) or increasing levels of PEG-induced water stress (–0.6, –0.8, and –1 MPa). Statistical significance was assessed using an unpaired t-test. Significance levels are indicated as follows: ns (p > 0.05), * (p ≤ 0.05), ** (p ≤ 0.01), *** (p ≤ 0.001), **** (p ≤ 0.0001).*

In order to determine whether the beneficial effect of CAP treatment on seed germination in water stress conditions could rely on an improved water uptake by the seeds, we investigated the changes of moisture content during imbibition on water and on a -0.8 MPa solution for seeds of variety B (**Figure 9**). Water stress decreased markedly water uptake by whole achenes but it was not significantly stimulated by CAP treatment.

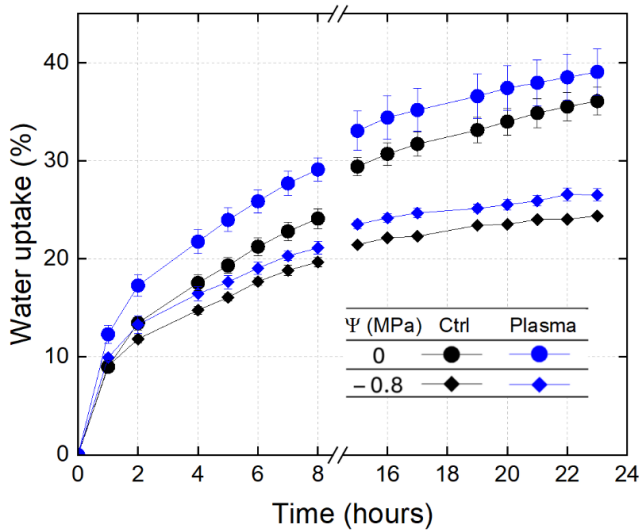

*Figure 9: Change in seed water content during imbibition of plasma-treated and control sunflower seeds from variety B under water and PEG -0.8 MPa conditions at 20 °C. Each point represents the mean water uptake calculated from 3 replicates of 20 seeds. Error bars indicate the standard error of the mean (SEM).*

### 3.4. Post-germinative effects of ambient air plasma on seedling growth

Since we established that CAP treatment potentially increased the germination rate of B and, more significantly, of A seeds under water stress, we then investigated whether the treatment could have an effect beyond germination, during the early stages of seedling development.

Seedling development was followed under well-watered and water stress conditions in a greenhouse maintaining a constant temperature of 21°C, which also allowed A seeds to germinate. Seeds were treated for 15 or for 30 min before being sown directly in the pots. In the absence of water stress both A and B seeds almost fully germinated in the greenhouse condition (soil and 21°C) (**Figure 10, Figure 11a, 11e**). As previously shown, water stress dramatically decreased germination of control A and B seeds (**Figure 10, Figure 11c, 11g**), which reached 20 and 40 % only, respectively. In A seeds, increasing durations of CAP treatment from 15 to 30 min allowed seeds to germinate to more than 60 % (**Figure 11c**) whereas in B seeds 15 min CAP treatment were sufficient to trigger germination rate higher than 80 % (**Figure 11g**).

We then evaluated the effect of CAP treatment on seedling growth. **Figure 10** shows the general appearance of seedlings that grew-up for 22 days with/without stress conditions. Supplemental Annex 1 shows the appearance of the seedlings at an earlier time point, *i.e.* 14 days. From these pictures it appears clearly that a 15 min CAP treatment was sufficient to stimulate seedling emergence in water stress conditions for A and B seeds.

Measurement of hypocotyl length after 19 days confirmed these observations. **Figure 11 b, 11d, 11f, 11h** show the frequency distribution of hypocotyl lengths according to treatments. In unstressed A seeds, 30 min CAP treatment slightly stimulated hypocotyl growth since most of the hypocotyls ranged in the 8-9 cm class, versus 7-8 cm for control and 15 min treated seeds (**Figure 11b**). This effect was not observed for B seeds (**Figure 11f**). For both A and B seeds CAP treatment induced a shift of hypocotyl mean lengths toward higher values, which was more important for A seeds (**Figure 11d**) than for B seeds (**Figure 11h**).

Finally, the effects of CAP treatment on both seed germination and seedling growth can be appreciated using the vigor I index which was calculated as the product of the average hypocotyl length (at day 19), by the germination rate (at day 19), divided by 100. In the case of A seeds, the CAP treatment increased the vigor index I from 0.9 (C) to 2.5 ($P_1$) or 4.4 ($P_2$) (**Figure 12a**). For the seeds in Batch B, this index increased from 2.3 (C) to 6.77 ($P_1$) or 6.25 ($P_2$) (**Figure 12b**). On the other hand, **Figure 12a, 12b** shows that in the absence of water stress, cold plasma had no effect, even significant.

*Figure 10: Photographs of sunflower seedlings, taken 22 days after seed sowing in pots placed in a greenhouse. Seedlings from variety A and variety B issued from untreated seeds (control) and from seeds subjected to plasma treatment in ambient air for 15 and 30 min.*

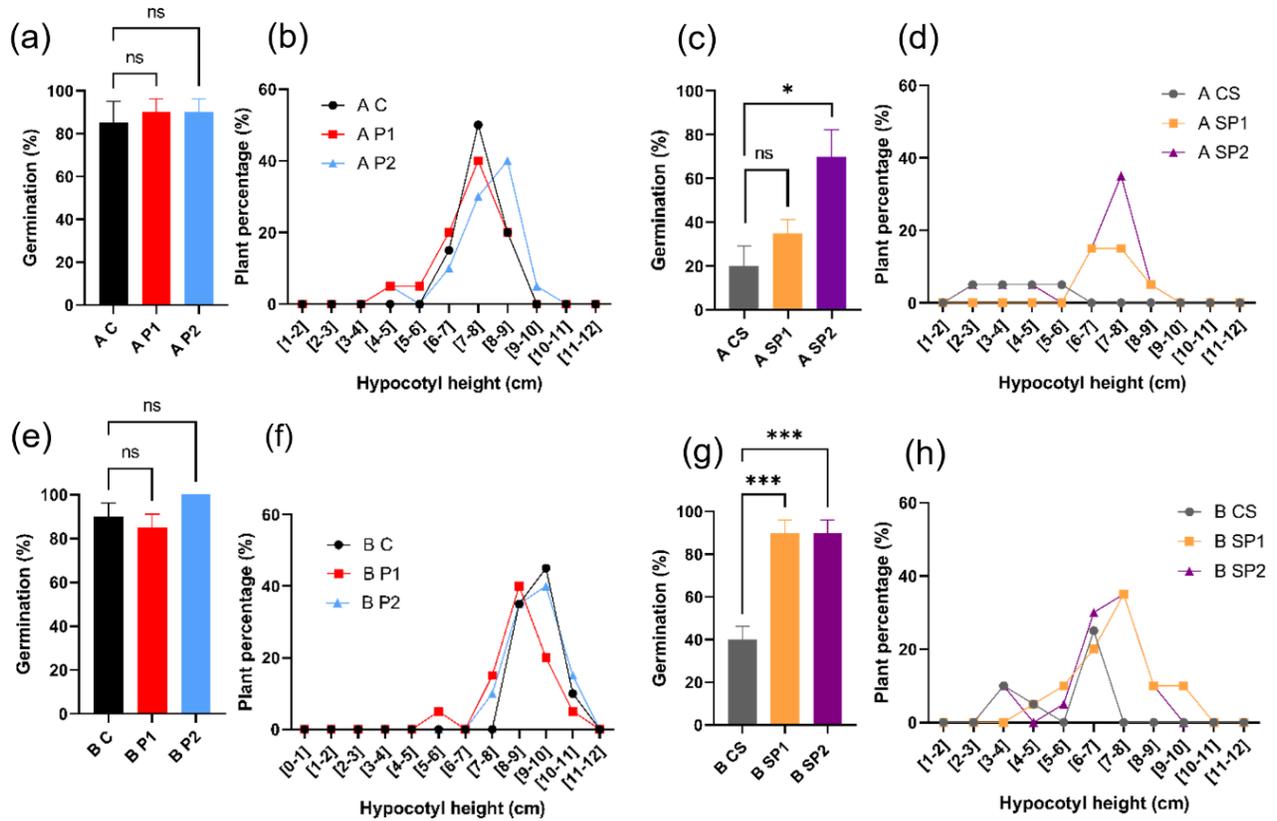

*Figure 11: Effect of plasma treatments on seed germination rate and hypocotyl growth for variety A (a, b, c, d) and variety B (e, f, g, h) under control (a, b, e, f) and water stress conditions (c, d, g, h) over a 19-day period. Hypocotyl length distributions (b, d, f, h) are shown as the percentage of seedlings falling within defined size intervals (e.g., [1–2] cm). C, untreated seeds, $P_1$ and $P_2$, seeds treated for 15 and 30 min by plasma, respectively. For histograms (a, c, e, g), each bar represents the mean percentage of germinated seeds per pot, calculated from five pots with four seeds each (20 seeds in total per condition). Statistical significance was assessed using an unpaired t-test. Significance levels are indicated as follows: ns (p > 0.05), * (p ≤ 0.05), ** (p ≤ 0.01), *** (p ≤ 0.001), **** (p ≤ 0.0001).*

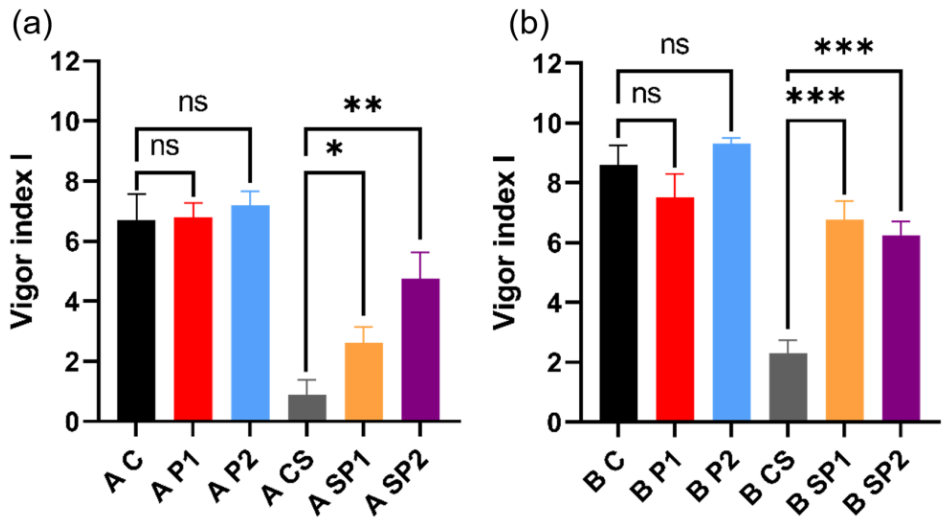

*Figure 12: Effect of plasma treatments on the vigor index I of sunflower seeds under control and water-stress conditions: (a) variety A, (b) variety B, over a 19-day period. The vigor index I corresponds to the product of the average germination rate and the average hypocotyl length, divided by 100. Each bar represents the mean vigor index I calculated for all pots per condition. Statistical significance was assessed using an unpaired t-test. Significance levels are indicated as follows: ns (p > 0.05), * (p ≤ 0.05), ** (p ≤ 0.01), *** (p ≤ 0.001), **** (p ≤ 0.0001).*

## 4. Discussion

The characterization of ambient air plasma generated using a dielectric barrier discharge (DBD) reactor revealed a complex, interdependent relationship between discharge parameters and the generation of reactive species. Notably, the production of strong oxidants such as ozone ($O_3$) in **Figure 6** was found to be highly sensitive to the electrode gap, in agreement with the findings of Kumar Shah et al.[50]. In addition, the physical filling state of the reactor (whether empty or packed with seeds) significantly influenced the yield and composition of gaseous reactive species, as previously reported by Molina et al.[51] and Judée et al.[52].

OES indicated that the plasma emission was dominated by the second positive system (SPS) of molecular nitrogen (**Figure 5**), which is characteristic of non-thermal atmospheric plasmas in air[46]. In contrast, short-lived oxygenated species such as atomic oxygen (O), hydroxyl radicals (OH), and nitric oxide (NO) were not detected. This absence is most likely due to rapid collisional quenching and energy transfer processes occurring under the specific discharge conditions used in this study, including atmospheric pressure, limited residence time, and frequent surface interactions[53]. Under such conditions, these transient species are either converted into more stable molecules before they can emit detectable photons or quenched via vibrational and rotational energy relaxation, rendering their optical signatures too weak to be captured by time-averaged OES measurements[54]. To overcome this limitation, advanced diagnostic techniques such as time-resolved OES, laser-induced fluorescence (LIF), or cavity ring-down spectroscopy (CRDS) could be employed to detect and quantify these highly reactive, short-lived species with greater sensitivity and temporal resolution[55].
Complementarily with OES, mass spectrometry enabled the detection of more stable species, particularly ozone, whose abundance significantly increased in the presence of seeds. Two factors could explain this trend: the dielectric nature of the seeds and their sharp geometrical structure.

Regarding seeds dielectric nature, it is worth reminding that seeds act as microscale capacitive elements within the discharge gap[46]. When densely packed, they modify the local electrical properties of the reactor by increasing the effective dielectric constant and contributing to heterogeneous charge storage[52]. This could result in altered electric field distributions and enhanced microdischarge activity, as observed in DBD systems packed with dielectric glass beads[56] and here in **Figure 4**. Such capacitive behavior likely facilitates a more intense and spatially distributed generation of ozone, providing a plausible explanation for the elevated $O_3$ concentrations detected in the presence of seeds.

Seeds of sunflower often present sharp tips and ridges. This morphology can induce local electric field enhancement at the apexes of the seed coat, a phenomenon well-documented in patterned dielectric barrier discharges (p-DBDs). Recent work by Berger et al., (2024)[57] demonstrated that dielectric pellets with sharper apexes (such as conical or inverse hemispherical shapes) exhibit distinct streamer dynamics compared to flat or rounded pellets. Specifically, sharper features lead to stronger local polarization, increased electron impact excitation rates, and higher streamer velocities, due to enhanced electric fields at the apex. By analogy, the natural topography of sunflower seeds may act as field-enhancing microstructures under DBD treatment, concentrating electric field and intensifying microdischarges in their vicinity. This localized enhancement likely promotes the formation and stabilization of reactive species, including ozone, via more efficient electron-driven reactions.

These physico-chemical properties have direct implications for the biological efficiency of the plasma treatment. Reactive nitrogen species such as $N_2^*$ and long-lived oxidants like $O_3$ are known to influence seed coat permeability and surface chemistry. $O_3$ may indeed facilitate oxidative degradation of inhibitory compounds or modify hydrophobic barriers on the seed surface[58]. Although OH and O radicals were not detected via OES, their possible transient presence at the plasma-seed interface cannot be excluded and may contribute to the activation of reactive oxygen species (ROS)-mediated signaling pathways involved in dormancy alleviation and water uptake enhancement.

Plasma treatments have been demonstrated to stimulate seed germination of a wide range of species (see Waskow et al. 2021 for review[59]) including in sunflower, but in this case using argon plasma[60, 61] or vacuum plasma[62]. In contrast to what is shown here, using a similar device (coaxial DBD reactor operated in air 10 min) but with different settings (sinusoidal voltage of 16 kV and amplitude at 50 Hz frequency) Florescu et al. (2023)[63] did not evidence any effect of plasma on germination of sunflower seeds, but they showed that the treatment could stimulate seedling growth. This demonstrates the importance of the plasma characteristics when designing a treatment to stimulate germination, as already suggested by Sarapirom and Yu (2021)[64]. Here we took advantage in phenotypic variability that exists between seeds of different hybrids of sunflower to explore the effect of plasma on both dormancy and germination under water stress. A short plasma treatment significantly stimulated germination of dormant seeds at 15°C (**Figure 7**), a temperature which allows expression of embryo dormancy[65]. Alleviation of embryo dormancy naturally occurs during dry storage (dry after-ripening), but may take several weeks or months[65]. Therefore, the ability to alleviate sunflower seed dormancy within a few minutes constitutes a striking advancement, with potential practical implications for the evaluation of seed quality and the acceleration of sunflower breeding programs. Sunflower seed dormancy alleviation during dry storage has been associated with non-enzymatic ROS generation[66]. Interestingly, it is also well known that plasma generates ROS, which might explain their beneficial effect on seed dormancy alleviation. Surprisingly the effect of CAP treatment on release of physiological dormancy is poorly documented but it has been evidenced in species as radish[67] or Arabidopsis[46]. Our results highlight that CAP is a promising treatment for releasing the dormancy of agroeconomically important species.

As previously demonstrated, the germination of sunflower seeds can be dramatically altered by water stress, but it also strongly relies on genotype[68]. As expected, dormant seeds were very sensitive to water stress and their germination rapidly became impossible with increasing PEG concentrations (**Figure 7**). However CAP treatment always stimulated germination of dormant seeds in suboptimal conditions of water availability, which resulted both from dormancy release, as previously shown, and from the improvement of germination in the presence of PEG. This was demonstrated using non-dormant seeds, which germinated quite well till -0.8MPa, but whose germination dropped down to 20 % at -1.0 MPa. In this later case, CAP treatment stimulated significantly germination, suggesting that plasma also improved tolerance to water stress at the germination stage. Interestingly CAP treatment significantly decreased $\Psi_{50}$ in a range that was quite similar to what can be obtained using seed priming. This effect has been shown for oilseed rape[41], wheat[44] or barley[69], and in most cases it was attributed to an improved water uptake by the seeds. This was not the case here, since we did not evidence a clear difference of water uptake between control and treated seeds in water stress conditions (**Figure 9**), which suggests that the better germination of CAP treated seeds relies on other mechanisms, that need to be explored.

Since stand establishment relies on both germination and seedling growth, we investigated the effects of CAP on the early stages of sunflower seedling development in the soil under water stress conditions. Besides stimulating seed germination under water constraints CAP also dramatically stimulated post-germinative seedling growth. This is clearly shown **Figure 11** where the length of the seedlings was generally 2-3 cm longer after 19 days of growth in a poorly watered soil. This demonstrates that an appropriate CAP treatment was highly beneficial for sunflower seedling emergence, since it stimulated both components of this process, *i.e.* the seed germination and the seedling growth, as demonstrated by the calculations of the vigor index I (**Figure 12**). In contrast, a decoupling of the effects of CAPs on germination and growth has often been reported, *i.e.* a positive effect on seedling growth without changes to the germination[70]. This discrepancy between effects on those two distinct developmental processes is not surprising since germination consists in cell elongation only whereas seedling development also requires active cell division through meristematic activity. Nevertheless, it is worth noting that a plasma treatment performed on dry seeds can last over time and modify seedling tolerance to water stress. It would be interesting to determine whether this treatment has also an effect on the long term; i.e. seed yield, and to better understand the mechanisms involved in this long-lasting effect.

# 5. Conclusion

Altogether our results demonstrate that CAP treatment is a promising technology to enhance crop emergence and resilience in changing environmental conditions. A key finding of this work is the critical importance of accurately adjusting the gap distance between electrodes in the plasma reactor. This highlights the need for precise engineering control in the design of plasma-based seed treatment systems, particularly for seeds with non-uniform or large size distributions. Future studies will have to determine if tolerance to other stresses during stand establishment, such as hypoxia or extreme temperatures, can also be improved by plasma. Further research should also focus on optimizing plasma treatment parameters, including power, duration, and gas composition, to maximize benefits for different crop species and environmental scenarios. Investigating the mechanisms of action of the plasma species, such as reactive oxygen and nitrogen species, in promoting seed germination, seedling growth, and stress tolerance is also crucial. Advancing our understanding of CAP technology should support sustainable crop production in the face of environmental challenges.

# 6. Acknowledgements

This work was supported by a PhD grant from Sorbonne Université (IPV programme) and received financial state aid as part of the PF2ABIOMEDE platform co-funded by « Région Ile-de-France » (Sesame, Ref. 16016309) and Sorbonne Université (technological platforms funding). The authors express their gratitude to the Institute of Environmental Transition and the Physics Department of Sorbonne University for the financial support provided to their research within the framework of the AMI project.

# 8. Annex

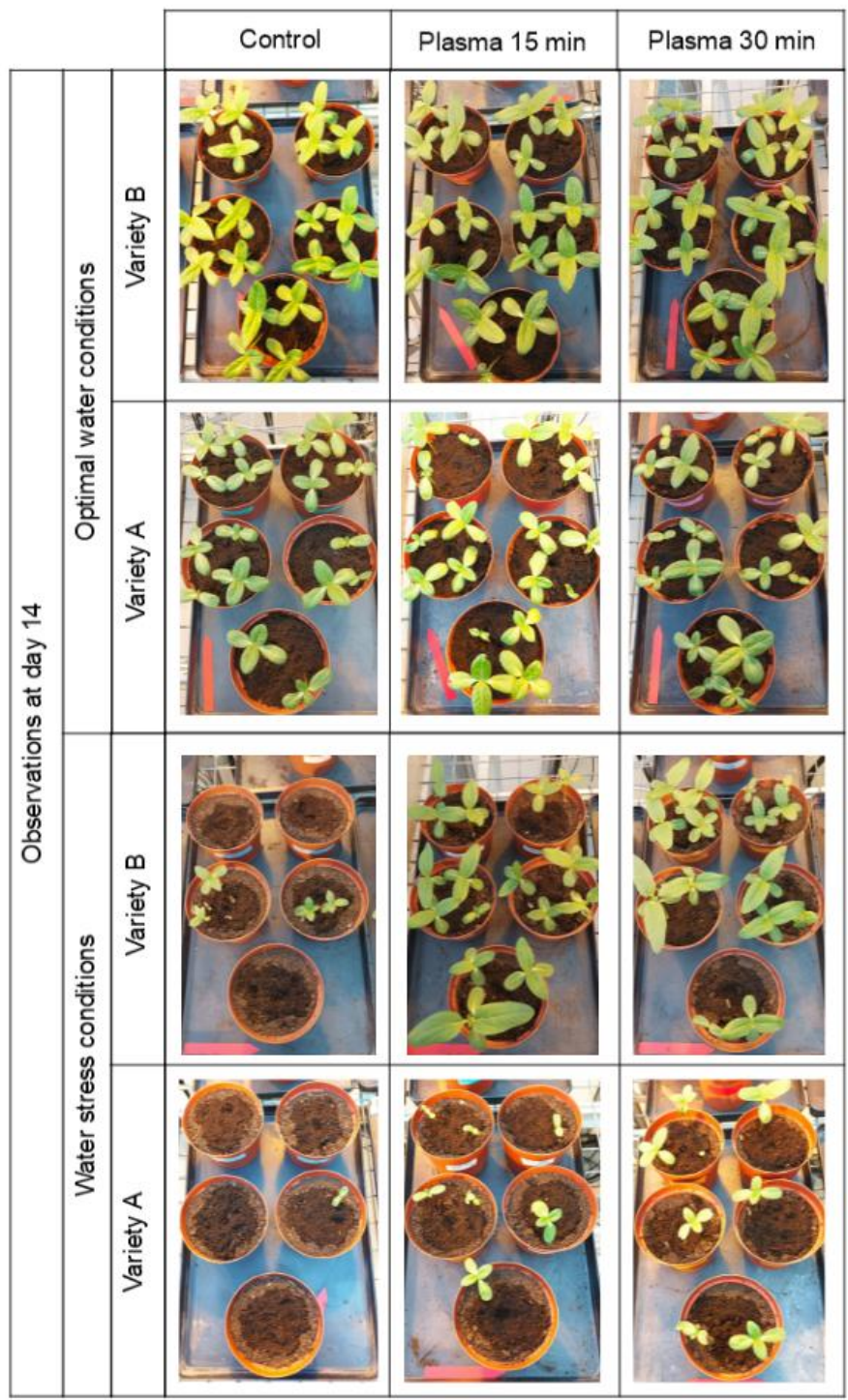

**Photographs of sunflower seedlings, taken 14 days after seed sowing in pots placed in a greenhouse. Seedlings from variety A and variety B issued from untreated seeds (control) and from seeds subjected to plasma treatment in ambient air for 15 and 30 min.**